# Polymyxin B-enriched exogenous lung surfactant: thermodynamics and structure

Nina Královič – Kanjaková[1$], Ali Asi Shirazi[1$], Lukáš Hubčík[1], Mária Klacsová[1], Atoosa Keshavarzi[1], Juan Carlos Martínez[2], Sophie Combet[3], José Teixeira[3], Daniela Uhríková[1#]

[1]*Department of Physical Chemistry of Drugs, Faculty of Pharmacy, Comenius University Bratislava, 832 32 Bratislava, Slovakia*

[2]*ALBA Synchrotron Light Source, Cerdanyola del Vallés, 08290 Barcelona, Spain*

[3]*Laboratoire Léon-Brillouin (LLB), UMR12 CEA, CNRS, Université Paris-Saclay, F-91191 Gif-sur-Yvette CEDEX, France.*

[#]Corresponding author, e-mail: uhrikova@fpharm.uniba.sk
[$]The authors contributed equally.

ORCID:  Daniela Uhríková    0000-0002-4397-1283

**Abstract:**

The use of exogenous pulmonary surfactant (EPS) to deliver other relevant drugs to the lung is a promising strategy for combined therapy. We evaluated the interaction of polymyxin B (PxB) with clinically used EPS, the poractant alfa Curosurf® (PSUR). The effect of PxB on the protein-free model system (MS) composed of four phospholipids (diC16:0PC/16:0-18:1PC/16:0-18:2PC/16:0-18:1PG) was examined in parallel to distinguish the specificity of the composition of PSUR. We used several experimental techniques (differential scanning calorimetry, small- and wide-angle X-ray scattering, small angle neutron scattering, fluorescence spectroscopy, and zeta potential) to characterize the binding of PxB to both EPS. Electrostatic interactions PxB - EPS are dominant. The results obtained support the concept of cationic PxB molecules lying on the surface of the PSUR bilayer, strengthening the multilamellar structure of the PSUR as derived from SAXS and SANS. A protein-free model MS mimics natural EPS well but was found to be less resistant to penetration of PxB into the lipid bilayer. PxB does not affect the gel-to-fluid phase transition temperature $T_m$ of PSUR, while $T_m$ increased ~ +2 °C in MS. The thickness of the lipid bilayer ($d_L$) of PSUR slightly decreases as a result of PxB binding, however, the hydrophobic tail of the PxB molecule does not penetrate the bilayer as derived from SANS data analysis and changes in lateral pressure monitored by excimer fluorescence at two depths of the hydrophobic region of the bilayer. Changes in $d_L$ of protein-free MS show a biphasic dependence on the adsorbed amount of PxB with a minimum close to the point of electroneutrality of the mixture. Our results do not discourage the concept of a combined treatment with PxB - enriched Curosurf®. However, the amount of PxB must be carefully assessed (less than 5 wt% relative to the mass of the surfactant) to avoid inversion of the surface charge of the membrane.

**Introduction**

Polymyxin B (PxB) is a naturally occurring cationic cyclic decapeptide isolated from *Paenibacillus polymyxa*[1]. The molecule of PxB (Supporting information, Fig. S1) is amphiphilic, composed of a cyclic heptapeptide head group and a linear lipophilic acyl chain. Five cationic diaminobutyric acid (Dab) residues secure electrostatic attraction to the negatively charged components of the lipid membrane. The lipophilic acyl tail forms 6-methyloctanoic acid (polymyxin B1) or 6-methylheptanoic acid (polymyxin B2)[2]. The cationic ring makes polymyxins soluble in aqueous environments, while the lipophilic chain facilitates insertion into bacterial membrane. PxB is highly bactericidal for Gram-negative bacteria due to its high affinity binding to bacterial endotoxin lipopolysaccharide (LPS), particularly its anionic lipid A phosphates. The proved nephrotoxicity and neurotoxicity restricted the clinical use of polymyxins[3]; however, the alarming global increase of antibiotic-resistant Gram-negative bacteria enforced their return, mainly as last-line therapy[4,5]. In addition to the urgent need for new antimicrobials effective against multidrug-resistant pathogens, various strategies to preserve the clinical efficacy of polymyxins and stop the development of polymyxin resistance are of interest. For example, drug repurposing approaches by combining polymyxins with nonantibiotics[6]; formulations and delivery systems to improve drug bioavailability and reduce its toxicity through targeted and controlled release[7,8]; rational molecular engineering based on structure-activity relationships[9–12], and others. The biophysical aspects of the effect of polymyxins on both LPS and lipid bilayers of various composition to understand the molecular mechanisms of the interaction were largely studied by using of many experimental and molecular simulation methods[10,13–19]. Interestingly, the therapeutic effectiveness of polymyxin B entrapped in extruded liposomes prepared from DPPC/Cholesterol=2:1 mol/mol was demonstrated in intratracheal administration to the lungs of rats infected with *P. aeruginosa*[20].

Endogenous pulmonary surfactant is a lipid-proteins mixture synthesized by type II alveolar cells, lining the inner surface of the lung. It reduces surface tension at the air-water interface, facilitating gas exchange and ensuring alveolar stability during respiration, and it is an important component of the innate host defense against invading pathogens[21–23]. Its inactivation, deficiency, or absence can result in life-threatening lung dysfunction. Intratracheal administration of exogenous pulmonary surfactant (EPS) (obtained from animals, for example porcine - Curosurf®; bovine -Survanta®) is currently standard therapy in neonatal intensive care, and has been tested in clinical trials for ventilated patients with Covid-19[24,25]. A more recent trend involves the idea that EPS could also be used as a delivery vehicle for pulmonary therapeutics and other compounds[26–28].

In this regard, the concept of PxB enriched exogenous pulmonary surfactant for the treatment or prophylaxis of Gram-negative pulmonary infections comes into focus. The addition of 2% PxB improves the surface activity of porcine EPS, Curosurf® at low concentration, increases its resistance to albumin inactivation, and reduces the incidence of pneumothorax in immature newborn rabbits undergoing prolonged ventilation[29]. PxB alone shows negative effects on alveolar type II cells, but not porcine surfactant or porcine surfactant enriched with PxB[30]. PxB was found to improve the resistance of modified porcine EPS to meconium *in vitro*, while antimicrobial function is maintained[31]. Prophylactic intratracheal treatment of neonatal rabbits with PxB/EPS prevents the growth of bacterial *E.coli*[32] and improves lung function in neonatal pneumonia of rabbits[33]. Encouraging results were reported in a similar model of pneumonia with polymyxin E, where the mixing of polymyxin with surfactant increased its bactericidal effect[34]. The authors assume a more efficient spreading of the mixture mediated by interactions between polymyxin E and surfactant. PxB improved the positive effect of EPS on inflammation and oxidative stress in mechanically ventilated rats with LPS- induced acute lung injury[35]. Calkovska et al.[36] proved the synergetic effect of surfactant therapy (Curosurf®) enriched with PxB in a double hit model of neonatal lung injury of immature rabbits. Independently, the authors evaluated changes in the surface activity of the surfactant exposed to smooth LPS without and with PxB using a captive bubble surfactometer. Indeed, our previous experiments[37] show that even ~ 1% by weight of LPS prevents the EPS (Curosurf®) from reaching the necessary low surface tension during area compression in a dynamic system that mimics the respiratory cycle. Small-angle X-ray scattering shows that LPS intercalates into the lipid bilayer of EPS and disturbs its lamellar structure by swelling. PxB acts as an inhibitor of LPS-induced structural disarrangement, restoring the original lamellar packing of EPS and its surface properties. The beneficial effect of PxB is attributed to its ability „to connect" negatively charged lipid vesicles. PxB exhibits an *in vitro* surface activity that is indistinguishable from that of analogous mixtures containing surfactant-associated specific protein SP-B[38]. Furthermore, PxB promotes the intermembrane transfer of phospholipids[19]. Therefore, there is a synergic effect of surfactant and PxB, leading to a possible combination therapy. However, the mutual interactions of EPS and PxB must be carefully assessed.

Here, we present results complementing the information on interactions of PxB with the clinically used exogenous pulmonary surfactant, poractant alfa Curosurf® (PSUR). Morphologically, PSUR is a mixture of unilamellar, oligo-, and multilamellar vesicles (ULV, OLV and MLV - "onion-like structure"). Illustrative microscope images of the PSUR non-homogenous suspension are shown in Fig. S2. The sample observed with polarized light (Fig. S2B) depicted "Maltese crosses", typical of multilamellar structures (as previously reported[37,39]). In the medical studies reported above, the PSUR suspension was mixed with the PxB solution (eventually, incubated for a defined time) prior to intratracheal instillation in animal models. We selected experimental methods for the study keeping in mind this application protocol. First, due to the natural origin of PSUR, the biophysical characteristics of PSUR itself were examined, along with repeated measurements using different batches of the nonexpired drug. Futhermore, the effect of PxB on the protein-free model system of pulmonary surfactant (MS) was examined in parallel with the aim of ascertaining the specificity of the EPS composition. The chosen model system composed of four synthetic phospholipids (diC16:0PC, 16:0-18:1PC, 16:0-18:2PC, 16:0-18:1PG; specified in the Material) was previously tested in animal models, and when completed with SP-B and SP-C analogues, has shown activity similar to that of PSUR[40]. We used several experimental techniques to characterize the PxB – EPS interaction: The binding of PxB to the EPS was monitored through changes in the surface charge of the lipid bilayer by measuring the zeta potential. Differential scanning calorimetry (DSC) was used to characterize the thermodynamic properties and assess the effect of PxB on the temperature of the gel- to liquid-crystalline (fluid) phase transition of both EPS. Structural changes in the lamellar packing were examined by small- and wide-angle X-ray scattering (SAXS/WAXS). A small- angle neutron scattering (SANS) measurements unraveled the effect of PxB on the lipid bilayer thickness of EPS and an aggregation of UVL into OLV. Finally, we aimed to elucidate the information about PxB localization within the EPS by monitoring the bilayer lateral pressure change using excimer fluorescence spectrometry.

**Material and Methods:**

*Material:*

Curosurf® (poractant alpha; Chiesi Farmaceutici, Parma, Italy) is a porcine pulmonary surfactant (we abbreviate as PSUR) provided at a concentration of 80 mg/mL. Polymyxin B sulfate salt (PxB) was purchased from Sigma-Aldrich (St. Louis, Missouri, USA). Synthetic lipids (DPPC, 1,2-dipalmitoyl-phosphatidylcholine (diC16:0PC); POPC, 1-palmitoyl-2-oleoyl-phosphatidylcholine (16:0-18:1PC); PLPC, 1-palmitoyl-2-linoleoyl-phosphatidylcholine (16:0-18:2PC); POPG, 1-palmitoyl-2-oleoyl-phosphatidylglycerol (16:0-18:1PG) (sodium salt)) were purchased from Avanti Polar Lipids (Alabaster, USA) and used without further purification. Fluorescence probes (Pyr4PC, 1,2-bis-pyrenebutanoyl-phosphatidylcholine; Pyr10PC, 1,2-bis-pyrenedecanoyl-phosphatidylcholine) were purchased from Thermo Fisher Scientific (Massachusetts, USA). Organic solvents, chloroform and methanol, of spectral purity, were purchased from Slavus (Bratislava, Slovak Republic). 150 mM NaCl stock hydration medium was prepared by dissolution of the appropriate amount of sodium chloride (Lachema, Brno, Czech Republic) either in MiliQ water (18.2 MΩ.cm, Millipore, Molsheim, France) or in heavy water (isotopic purity 99.9% $D_2O$; Merck, Germany).

*Sample preparation:*

The proteolipidic mixture (PL) of PSUR was diluted by hydration medium at the required concentrations. The pulmonary surfactant model system (MS) was prepared from synthetic phospholipids diluted in a mixture of chloroform and methanol (3:1 by volume) and mixed to obtain DPPC:POPC:PLPC:POPG=50:24:16:10 wt%. The organic solvent was evaporated under a stream of nitrogen gas, and its remaining residues were removed under vacuum. The dried lipid films were hydrated by adding 150 mM NaCl and homogenized by vigorous vortex mixing and by freezing/thawing cycles (-50 °C/50 °C), repeated at least four times. For SANS measurements, unilamellar vesicles were prepared as described hereafter. PxB was hydrated with 150 mM NaCl and homogenized by vortexing. The PxB/EPS (PSUR or MS) mixtures were prepared by mixing appropriate volumes of hydrated compounds and gently shaking incubated at 45 °C for 30 min up to 1 hour. pH changed in the range 5.6±0.3. The amount of PxB in the sample is reported in weight percent (wt%) respect to the mass of the EPS (PSUR or MS).

For SANS measurements, PSUR suspension was dried above $P_2O_5$ desiccant in a $N_2$ atmosphere at ~ 7 °C. The dried PSUR was rehydrated with $D_2O$ to a 2 wt% solution. For fluorescence measurements, a portion of dried EPS (PSUR, MS) was dissolved in a mixture of chloroform and methanol (3:1 by volume) and mixed with fluorescence probes in the ratio of 1500:1 mol/mol. The organic solvent was evaporated under a stream of nitrogen gas, and its remaining residues were removed under vacuum. The dried EPS films were hydrated by the addition of appropriate volumes of PxB solution and filled to the desired volume with hydration medium. The mixtures were homogenized by repeated vortex mixing and mild sonication and stored 24 hours in the refrigerator for equilibration[41].

*Methods:*

*Zeta Potential Measurements*

The zeta potential was measured by electrophoretic light scattering of the particles using a Litesizer 500 particle size analyzer (Anton Paar, Austria). Multilamellar vesicles at a concentration of 3 mg/mL in a 150 mM of NaCl were measured at 37 °C. Polycarbonate disposable Omega cuvettes for the zeta potential (Anton Paar, Austria) with volume 900 μL were used for the measurement. The zeta potential was evaluated with Kalliope software using the Smoluchowski model. For the calculation, the viscosity 0.707346 mPa.s, the dielectric constant 72.415 F/m, and a refractive index of 1.330434 were assumed. The reported average values of the zeta potential and standard deviations are obtained from 100 measurements for each sample.

*Differential scanning calorimetry (DSC)*

Lipid mixtures at a concentration of 3 mg/mL were used for DSC measurements. PxB/EPS samples were prepared as described above and stored in a refrigerator for 24 hours. Before measurement, the sample was degassed using a degassing station (TA Instruments, USA) for 15 min at temperature 4 °C and pressure 230 Torr. A Nano DSC calorimeter (TA Instruments, USA) was used for the measurement. Three scans (heating – cooling - heating) were performed in the range of 0 – 60 °C with a heating rate of 1 °C/min at a pressure of 3 atm and a pre-equilibrium halt of 1000 s for each sample. An example of the DSC heating/cooling profile is shown in Fig. S3A. Data were treated using the Origin (Pro) program (Northampton, MA, USA). Calorimetric enthalpies ($\Delta H$) were analyzed using a standard integration procedure of the areas under the peaks after correction of the baseline and normalization to the mass of the sample. The temperature of the main phase transition ($Tm$) was derived from the maximum of the peak. Reported values represent the mean $Tm$ of the first and third scans (heating profiles).

*X-ray scattering (SAXS/WAXS)*

Lipid mixtures at a concentration of 10 mg/mL were used for the X-ray experiment. Before measurement, samples were centrifuged at 13,000 rpm for 2 min using a Minispin centrifuge (Eppendorf, Germany). The sediment was transferred to thin-walled borosilicate glass capillaries (WJM-Glas, Berlin, Germany) of 1.5mm diameter and closed with a plasticine.

Small-angle (SAXS) and wide-angle (WAXS) synchrotron radiation scattering experiments were performed on the BL11-NCD-SWEET beamline, ALBA Synchrotron, Barcelona (Spain) using linearly polarized radiation with a wavelength $\lambda = 0.1$ nm. The capillary was placed vertically in a Linkam stage, which provided temperature control. Samples were measured at selected temperatures and / or by temperature scans. The heating program was as follows: 3 min incubation of a sample at fixed temperature, an exposure time of 0.5 or 1 s, and heating to the next temperature. The heating rate during the temperature scans was 1 °C/min and the patterns were recorded with an exposure time of 0.5 s every minute. SAXS data were detected on a Pilatus 3S 1M detector calibrated using silver behenate[42]. WAXS data were detected on an LX255HS Rayonix detector calibrated with $Cr_2O_3$ (certificate SRM 674b, NIST, USA). The 2D scattering patterns were azimuthally integrated into 1D data using the pyFAI python library[43]. The raw data were normalized to the intensity of the incident beam. The diffraction maxima were derived by fitting each peak with the Lorentzian function and linear background using the Peakfit software. The repeat distance $d$ of the lamellar phase $d= n.2\pi/q_h$, ($n =1, 2,…$) where $q$ is the position of the maximum of the Bragg peak of $n$ – order in the SAXS pattern. The repeat distance $d$ and its uncertainty were derived from maxima of all observed peaks belonging to one lamellar phase. The repeat distance $d=d_L+d_W$ involves the thickness of the lipid bilayer $d_L$ and the thickness of the water layer $d_W$, localized between adjacent lamellae.

Structural changes induced by PxB were followed in the fluid $L_\alpha$ phase, for selected samples temperature scans are shown in Supporting information (SAXS/WAXS). WAXS patterns of all the mixtures studied exhibited a broad peak with a maximum at ~ 14 nm$^{-1}$, characteristic of lipids with liquid-like acyl chains packed in a quasi-hexagonal lattice.

*A small angle neutron scattering (SANS)*

Unilamellar vesicles were prepared by extrusion of hydrated EPS through a polycarbonate filter (Nuclepore Plesanton, Canada) with pores of 100 nm diameter, using the LiposoFast Basic extruder (Avestin, Ottawa, Canda) fitted with two gas-tight Hamilton syringes (Hamilton, Reno, USA) as described[44]. Samples were subjected to 51 passes through the filter at a temperature ~ 45 – 50 °C. The resulting solution showed only a marginal opalescence, which is typical for the dispersion of unilamellar vesicles (ULV). Dynamic light scattering (Litesizer 500, Anton Paar, Austria) has shown ULVs of diameter 110±20 nm with polydispersity ~ 15 % (at 40 °C). The ULVs were incubated with PxB at 45 °C for 30 min up to 1 hour. A lipid dispersion of 1 wt% filled in 2 mm thick quartz cells (Hellma, Germany) was used for the SANS measurement.

Neutron scattering experiments were performed on the PAXY spectrometer of the Orphée reactor (Laboratoire Léon Brillouin, Saclay, France). The scattered intensity *I(q)*, is measured as a function of the momentum transfer, *q*, which depends on the wavelength, $\lambda$, of the incident neutron beam and the scattering angle $2\theta$; $q=4\pi.sin(\theta)/\lambda$. The temperature was set to 37±0.1 °C and acquisition time for each sample was 30 min. The samples were measured at the sample-to-detector distance of 3.195 m, and the neutron wavelength $\lambda = 5$ Å (resolution $\Delta\lambda/\lambda = 10\%$). This setup covers an optimal range of $q =0.1 – 2.5$ nm$^{-1}$ required to obtain the value of the thickness of the lipid bilayer ($d_L$) in ULV[45] at one acquisition time. The normalized SANS intensity, *I(q)*, was obtained using the *Pasinet* software provided by LLB. Spectra were corrected for an incoherent background. The thickness of the lipid bilayer ($d_L$) was determined by fitting the experimentally obtained *I(q)* vs. *q* using the SasView software[46] using two built-in models. A model of an unilamellar vesicle was used for the analysis of PSUR and MS scattering curves (SANS analysis, Fig. S9). The SANS patterns of PxB – EPS (PSUR and MS) were fitted by using a paracrystal lamellar model[47] (SANS analysis, Supporting information). The ideal model scattering curve was smeared by the instrumental resolution function. For the fitting, the neutron scattering length density (NSLD) of PSUR, $\rho_{PSUR}$, was determined experimentally from the matching point of the scattering densities of the PSUR and $D_2O/H_2O$ mixtures (Supporting information, Fig. S10). The average NSLD of MS, $\rho_{MS} = 0.273 \cdot 10^{-6}$ Å$^{-2}$ was determined with respect to the composition and molecular volumes of lipids[48–50].

*Lateral pressure measurements*

Lateral pressure was monitored at two depths of the PL bilayer, using excimer fluorescent probes with pyrene moieties located in the fourth carbon (Pyr4PC) and the tenth carbon (Pyr10PC) of the acyl chain[51]. These probes randomly mix in fluid and gel bilayers and are able to mimic the physical properties of nonfluorescent PCs in the bilayer[52]. Measurements were made using a FluoroMax-4 spectrofluorimeter (HORIBA Jobin Yvon, France) in a quartz cuvette with an optical path length of 1 cm. Measurements were made at a temperature of 45.00 ± 0.01 °C, maintained by a Peltier thermocouple drive. The samples were continuously stirred during measurement. The samples were excited at a wavelength of $\lambda_{ex}$ = 345 nm and the emission spectrum was collected in the region of $\lambda_{em}$ = 360–650 nm with an increase step of 0.5 nm. The excitation and emission bandwidths were 5 and 3 nm, respectively, and the integration time for each wavelength was 0.1 s. The fluctuation of the fluorescence intensity over time was monitored at two emission wavelengths $\lambda_{em}^{M}$ = 376 nm and $\lambda_{em}^{E}$ = 475 nm, corresponding to the emission of the monomer and excimer form of pyrene, respectively[53]. The ratio of signal detector (sample) to reference detector (lamp), S1c/R1c, corrected for the wavelength- and time-dependent spectral response, was monitored at both emission wavelengths. The lateral pressure η within the bilayer plane is given by the ratio of excimer to monomer intensities[54], $\eta = I^E/I^M$, where $I^E$ and $I^M$ denote the averaged value of excimer and monomer emission intensity, calculated from the time-based measurement.

**Results and discussion**

The structure and function of pulmonary surfactant are interconnected. The complexity of PS, the role of individual components, its structure and function have been largely discussed and excellently reviewed elsewhere[21,23,55–58]. The inactivation of endogenous pulmonary surfactant results in severe pulmonary diseases that are therapeutically solved by intratracheal administration of exogenous pulmonary surfactant (EPS) from animal sources. The clinically used surfactant replacement, Curosurf® (PSUR), is a proteolipid extract prepared from whole minced porcine lung tissue. Up to 78 wt% of all lipids (~ 99 wt%) of PSUR are phosphatidylcholines (PC). DPPC represents ~ 50 wt% of PC. Negatively charged lipids (PG, PI) can reach up to 10 wt%. Surfactant associated specific proteins (SP-B and SP-C) represent ~ 1 – 2 wt%. (For the composition, see Table S1, Supporting information and/or Oseliero Filho et al.[59]). SP-B (8.7 kDa) and SP-C (4.2 kDa) are small hydrophobic proteins. SP-B has 79 residues with a net positive charge +7. SP-C is a very hydrophobic protein; it has 35 residues and a net positive charge +3[60]. Even in small amounts, proteins together with negatively charged lipids (PG, PI) contribute to the surface charge of the PSUR. Zeta potential measurements show the PSUR as a suspension with a negative surface charge under physiological conditions[61]. We found zeta potential $\xi$ = -9.4±1.0 mV (at 37 °C). Electrophoresis does not reveal the direct surface potential, but the electric potential in the shear plane (~ 0.2 nm), i.e. the zeta potential $\xi$. Its value reflects the surface potential of the vesicles in solution[62]. The protein-free model system of pulmonary surfactant (MS) composed of a mixture of four synthetic lipids (DPPC, POPC, PLPC, and POPG) shows a zeta potential close to that obtained for natural EPS, $\xi$ = -13.6±1.0 mV at 37 °C.

The interaction between cationic PxB and negatively charged lipid mixtures of PSUR and MS was monitored by measuring the zeta potential of particles. Fig. 1 shows the change of the zeta potential $\xi$ of PSUR and MS vesicles incubated with PxB at 37 °C. Both lipid mixtures show similar dependencies with increasing PxB. Binding of cationic PxB molecules reduces the negative charge of the EPS surface, and $\xi$ reaches the electroneutrality at ~ 5 – 7 wt % of PxB. The zeta potential $\xi$ ~ 3.5 – 5 mV in mixtures with the content of PxB ≥ 10 wt% indicates positive charge of the lipid surface.

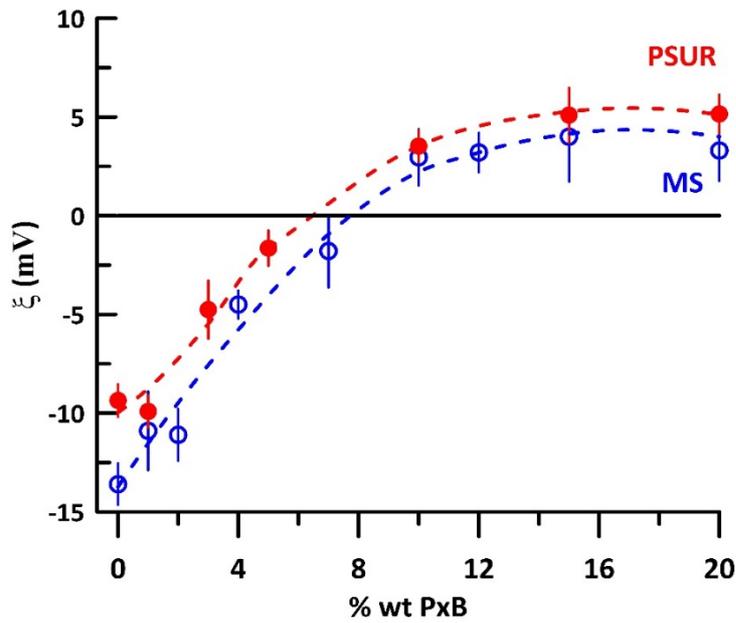

Fig. 1 The zeta potential of PSUR (red circles) and MS (blue circles) in the presence of PxB at 37 °C.

*Thermodynamics*

Fig. 2 summarizes the DSC results. Fig. 2A shows typical endotherms of PSUR (red curves) and MS (blue curves) incubated with PxB (as described in Materials and Methods). The full lines indicate the temperature of the main phase transition, $T_m$, of the EPS (PSUR and MS) without PxB.

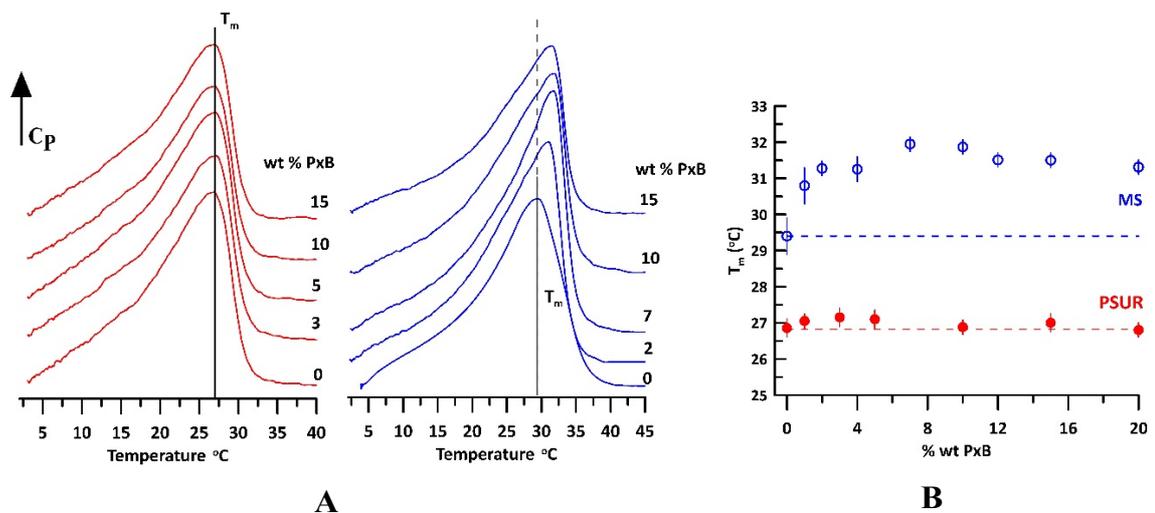

Fig. 2A DSC profiles of the PSUR (red) and protein-free MS model system (blue) in the presence of polymyxin B (PxB). The full lines denote the temperature of the phase transition $T_m$ from gel to fluid state of the EPS without the antibiotics.

Fig. 2B The temperature of the phase transition *Tm* from gel to fluid state of PSUR (red) and MS (blue) in the presence of PxB (wt%). The dashed lines represent the value of *Tm* of PSUR and MS, respectively, without PxB.

A broad asymmetric endothermic peak (with width, FWHM ~ 12 °C) denotes a complex transition from gel ($L_\beta$) to fluid phase (liquid-crystalline $L_\alpha$) with low cooperativity, reflecting the composite lipid composition with large differences in the temperatures of gel-fluid phase transitions (*Tm*) for individual species. Zwitterionic DPPC, the main surface active lipid component in the native pulmonary surfactant and also in EPS[63], shows the main phase transition at *Tm* ~ 41 °C[64]. The presence of one double bond in the lipid acyl chain results in a drop in *Tm* (for example, -5.3 °C for POPG (16:0-18:1PG), -2.5±2.4 °C for POPC (16:0-18:1PC), and PLPC (16:0-18:2PC) has *Tm* -18 °C [65–67]). The size of vesicles (MLV or ULV) can affect thermodynamic parameters[65]. We did not detect significant differences in the thermograms of the PSUR suspension, or the ULVs prepared by extrusion as described in Materials and Methods (Fig. S3B). Lipid mixtures were prepared in excess of water; thus, the unsaturated lipids (*e.g.,* POPC, PLPC, POPG) are in the fluid state in the scanned temperature range of DSC measurements. The asymmetric peak of DSC patterns is characteristic of a lipid membrane structurally composed of fluid-gel domains, as reported for mixtures of DPPC/unsaturated lipids[68–70]. The presence of domains that spread laterally in the lipid matrix of native pulmonary surfactant and/or therapeutically used EPS was proved[55,71–73].

We determined phase transition temperatures ranging from 26.8 to 28.2 °C with the mean value *Tm* = 27.3±0.5 °C from the 1st and 3rd scan (heating-cooling-heating regime) of six independent samples of PSUR (different batches, not expired drug, MLV or ULV) using the same experimental protocol. The obtained data correspond well to recently reported *Tm* ranges between 27.0 and 29.5 °C[59,74–77]. The thermogram of protein-free MS shows the maximum at *Tm* = 29.4±0.5 °C. The enthalpies derived by integration of the excess of heat capacity ($c_p$) profile above a linear background, delimited by measuring range (3 - 50 °C) yield *ΔH* = 21.1±3.0 J/g and 24.0±1.0 J/g for PSUR and MS, respectively. The values are roughly one-half of the enthalpy of DPPC[65,78] and correspond well to the composition of EPS. Protein-free MS mimics well the thermal behavior and thermodynamic properties of PSUR. The similarity of thermograms of both systems indicate that a small content of proteins in PSUR does not affect the DSC profile significantly.

Clearly, the binding of PxB to PSUR does not affect its phase transition temperature ($L_\beta \rightarrow L_\alpha$). Indeed, the *Tm* values derived from the endotherms (Fig. 2B, red symbols) are practically independent of the content of PxB. The dashed red line in Fig. 2B indicates the *Tm* of the PSUR mixture, $Tm=26.8$ (±0.3) °C. One may detect a slight increase in *Tm* in mixtures with 1 – 5 wt% of PxB. However, the change remains under ~ 0.3 °C. The obtained enthalpies fluctuate in the range of accuracy ΔH of the PSUR itself. The main phase transition from gel to fluid phase ($L_\beta \rightarrow L_\alpha$) is endothermic process related to the "melting" of the acyl chains of the bilayer[54]. To fully understand structural changes and thermodynamics of PxB – EPS (PSUR and MS), we review briefly the temperature induced polymorphism of the DPPC/water mixtures in the Supporting material, Fig. S4. WAXS reflects the organization of acyl chains in the hydrophobic region of the bilayer, and data collected in a scan with increasing temperature (so-called temperature scan) provide information about phase transitions. Figures S5 and S6 show such a temperature scan in the range 20 – 44 °C for PSUR without and with 3 wt% of PxB, respectively. We derived $Tm = 28.0 \pm 0.5$ °C from the temperature dependence of the first derivative $f(t) = dq/dT$, where $q$ is the position of the maximum derived from WAXS at temperature $T$. The addition of 3 wt% of PxB to PSUR does not affect *Tm* (Fig. S7). The values obtained correspond to the *Tm* obtained by DSC. The data obtained from both experimental methods indicate that PxB molecules do not affect the melting of the acyl chains of PSUR. The endotherms of the protein-free MS mix composed of four lipids (blue curves, Fig. 2A) indicate an increase in *Tm* due to binding of PxB (Fig. 2B, blue circles). *Tm* increases with PxB, from $Tm = 29.4 \pm 0.5$ °C for pure MS to ~ 31.8 °C in the mixture of PxB/MS with 7-10 wt% of PxB. Thus, PxB molecules impede trans-gauche isomerization and stimulate tighter packing of lipid acyl chains in the hydrophobic part of the protein-free MS bilayer in comparison to that of PSUR. Therefore, the obtained results reflect different binding of PxB molecules in the two studied systems, PSUR and protein-free MS.

For clarity, next study is focused on the effect of PxB on both EPS systems in the fluid state ($L_\alpha$ phase), close to physiological conditions (T ~ 37 - 45 °C).

*Structural study*
*Role of electrostatic interactions (SAXS and SANS)*

Generally, the repeat distance *d* of fully hydrated zwitterionic phosphatidylcholines (DPPC, POPC, PLPC and others) in the $L_a$ state does not exceed ~6.5–6.7 nm. From a thermodynamic point of view, the lipid of one chemical species hydrated with an excess of water forms a two-phase system composed of lipid lamellae separated by the layer of water (of thickness $d_W$) and the bulk water. The thickness of the water layer, $d_W \sim 1.8 - 2$ nm for PCs, is the result of a balance between repulsive interbilayer interactions (steric, hydration, and fluctuations) and attractive van der Waals forces[79]. The presence of uncompensated charges at the bilayer overcomes the attractive van der Waals forces, and the distance between bilayers increases. In the extremity, MLV can fall into OLV or ULV due to repulsive electrostatic force[80]. The increase in water thickness ($d_W$) between lamellae and their fluctuation causes a disorder in the relative positions of the unit cells (long-range disorder), resulting in a broadening of the diffraction peaks, and the magnitude of this effect increases with increasing diffraction order, and with temperature[79,81].

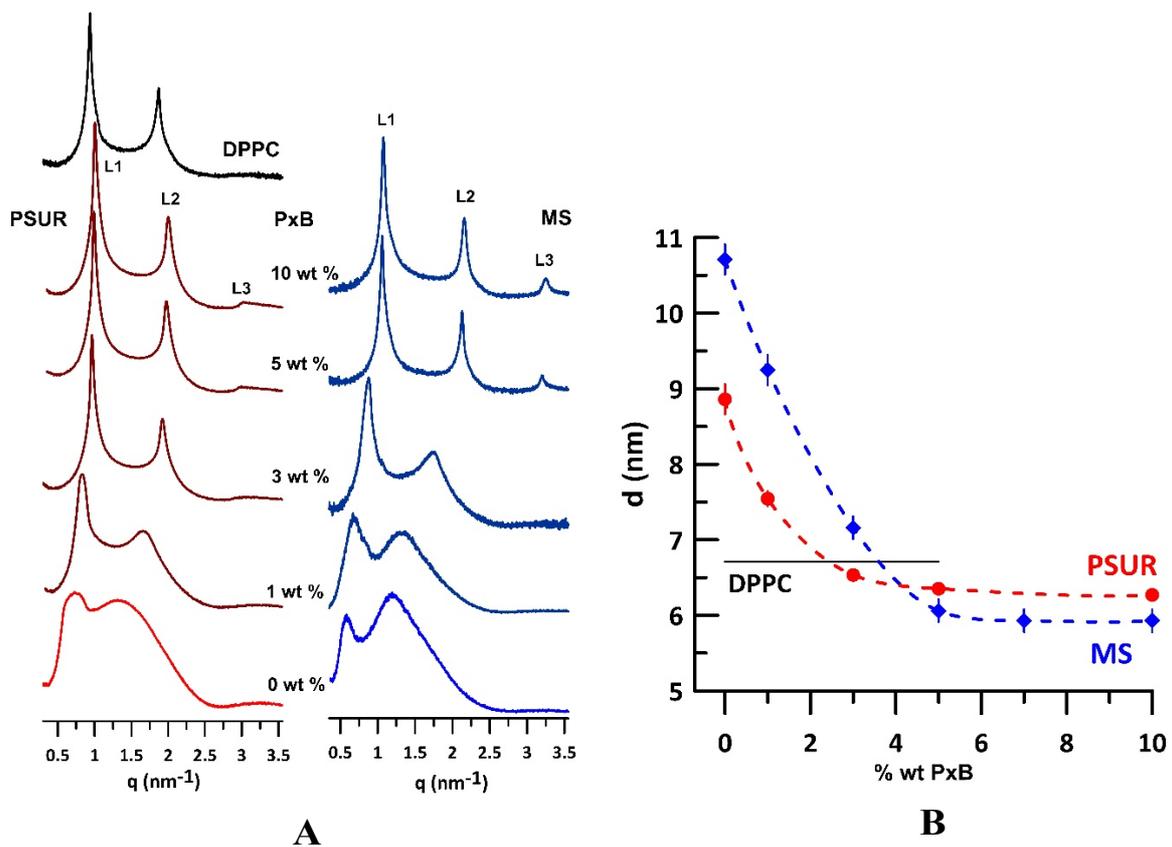

A B

Fig. 3A SAXS patterns of PSUR and MS without and with polymyxin B (PxB) in a fluid $L_\alpha$ phase (40 °C), and DPPC at 45 °C. The intensity is in logarithmic scale, and individual patterns are shifted along the y-axis for clarity.

Fig. 3B The repeat distance $d$ of a lamellar phase of PSUR (red circles) and protein - free MS (blue diamonds) in the presence of PxB (wt%) at 40 °C; full black line indicates the repeat distance $d$ of DPPC in $L_\alpha$ phase at 45 °C. Error bars include the uncertainties

of $d$ evaluated from fitting of SAXS peaks, and from the evaluation of patterns of two independent samples.

Figure 3A shows SAXS patterns of EPS (PSUR and MS) at 40 °C. Both systems are in fluid $L_\alpha$ phase and the WAXS pattern of all measured samples exhibited one broad peak with the maximum at q ~ 14 nm$^{-1}$, characteristic of liquid-like acyl chains of phospholipids, as shown in Figs. S4 and S5 (at $T >Tm$). Two broad peaks, $L(h)$, where $h$ =1, 2 is the Miller index, were attributed to a lamellar phase of swollen MLV derived from negatively charged bilayers (zeta potential ξ ~ - 10 mV). A detailed inspection of the EPS pattern revealed the presence of a broad peak of low intensity at $q$ ~ 1 – 1.2 nm$^{-1}$ that resulted in an increase in the baseline between the two peaks (shown in Figs. S5 and S6). We attribute this peak to a form factor of ULVs and/or OLVs present in EPS. We determined the repeat distance $d$ = 8.86±0.20 nm and 10.70±0.20 nm for PSUR and MS at 40 °C, respectively. Due to the natural origin of PSUR, its composition may be different. From our measurements, the PSUR repeat distances obtained for samples from different batches of the drug ranged: $d$ ~ 8.0 – 9.4 nm at 40 °C (and $d$ ~ 10.3 – 11.1 nm at 20 °C). The repeat distance $d$ decreases with increasing temperature, although, in the range of physiological temperatures (T ~ 35 – 45 °C) its change does not exceed ~ 0.4 nm as derived from the PSUR and MS temperature scan, respectively (Fig. S8). The repeat distance, $d$ ~ 9 nm, for PSUR at 40 °C is consistent with the previously reported $d$ ~ 8.3 – 9 nm for fully hydrated natural EPS (porcine, bovine) in the fluid state (~ 40 °C)[39,59,82], while smaller values $d$ ~ 7.2 – 7.5 nm were found in mixtures under reduced hydration[83,84].

PxB binding compensates for the negative surface charge of EPS as confirmed by zeta potential measurements (Fig. 1). The SAXS patterns in Fig. 3A show a significant improvement in the positional order of elementary cells (lamellae) when cationic antibiotic is added to negative charged EPS: we observe up to three peaks of $L_\alpha$ phase, significant decrease in their width, and their shift to higher $q$ values with increasing PxB content. EPS mixtures (PSUR, MS) enriched with ≥ 3 wt% of PxB form a well-ordered structure comparable to those detected in fully hydrated zwitterionic phosphatidylcholines, such as DPPC (the upper pattern in Fig.3A, left panel) with $d$ ~ 6.7 nm (marked by a solid line in Fig. 3B). Figure 3B shows the repeat distance, $d$, with increasing PxB content (in wt%) at 40 °C. In PSUR, the binding of PxB reduces $d$ from 8.86±0.10 nm, to 6.35±0.02 nm for 5 wt% of PxB in the mixture (full red symbols). Further addition of antibiotic yielded negligible variations of $d$ (< 0.10 nm). In protein-free MS, $d$ falls sharply from 10.70±0.20 nm to ~ 6.00 ± 0.10 nm for PxB > 3 wt% (blue diamonds).

Positively charged Dab residues of PxB play a dominant role in the interaction, targeting negatively charged moieties on phospholipid molecules (phosphate groups, carboxyl groups, and glycerol) and thus screen the negative surface charge of lipid bilayers that in turn facilitates a closer approach of lamellae reaching the level observed for zwitterionic lipids. Molecular dynamic simulations indicate even the formation of a complicated interaction network of the polar cyclic ring of polymyxins with the head groups of phospholipids[9,14,85]. Our data show a lamellar structure, reaching in the $L_\alpha$ phase (at 40 °C) a practically steady $d$ ~ 6.3 nm at the content of PxB ≥ 5 wt% in the concentration range studied (Fig. 3B). The zeta potential measurements (Fig. 1) indicate the electroneutrality of PxB -EPS (PSUR, MS) at ~ 5 – 7 wt% of PxB, which gives the stoichiometry of 1 molecule of PxB to 30 – 21 molecules of EPS (based on the composition). In our previous study, we have shown the ability of PxB to restore the original lamellar structure of the swollen PSUR up to $d$ ~ 12.6 nm when "infected" by 10 wt % of LPS[37]. The PxB molecule with a polar cyclic ring and five cationic Dab residues has "a large" polar area[13], A = 6.464 nm². The average area per molecule of the lipid forming MS represents A = 0.64 ±0.2 nm² based on the composition and data available from structural studies[48,86,87]. Therefore, the polar part of PxB can theoretically overlap the surface area up to ~ 10 molecules of lipids of MS, which is consistent with the concept of the PxB – lipid headgroup network as proposed by molecular dynamic simulation.

Our SAXS experiments did not discriminate the effects of PxB on the thicknesses of the lipid bilayer ($d_L$) and the water layer ($d_W$). The amphiphilic character of the PxB molecule predestines its insertion into the lipid bilayer, eventually causing its thinning, as proposed in the Shai-Matsuzaki-Huang model for the antimicrobial mechanism of peptides[88–90].

Morphologically, Curosurf® (PSUR) is a suspension of multi-, oligo-, and unilamellar vesicles. Molecules of PxB added to the PSUR thus interact with each of the structures. The presented SAXS data indicate rather strong effects, supporting an aggregation of lipids into MLV as a result of PxB. To clarify the effect of PxB on UVLs, we performed a small- angle neutron scattering (SANS) experiment taking advantage of the contrast between the coherent neutron scattering length densities (NSLD) of EPS lipids and solvent ($D_2O$). Unilamellar vesicles EPS (PSUR, MS) were prepared by extrusion, and analysis of SANS curves allowed to assess the changes in the thickness of the lipid bilayer ($d_L$) caused by PxB and an aggregation of ULVs.

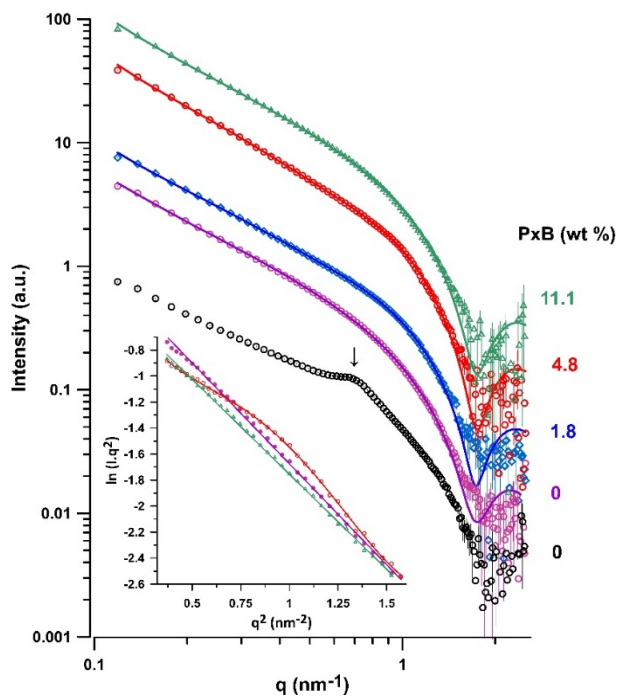

Fig. 4A SANS curves of a suspension of PSUR (black circles); unilamellar vesicles of PSUR (violet circles) and UVLs of PSUR incubated with PxB at 37 °C. The full lines represent the best-fit results using the model of polydisperse unilamellar vesicles for PSUR data and the paracrystal lamellar model for PxB – PSUR mixtures. Curves are shifted along the y-axis for clarity. Inset: Kratky-Porod plot, ln $I(q).q^2$ versus $q^2$, of PSUR (violet circles) and PSUR in the presence of 4.8 wt% (red symbols) and 11.1 wt% (green symbols) of PxB (at 37 °C).

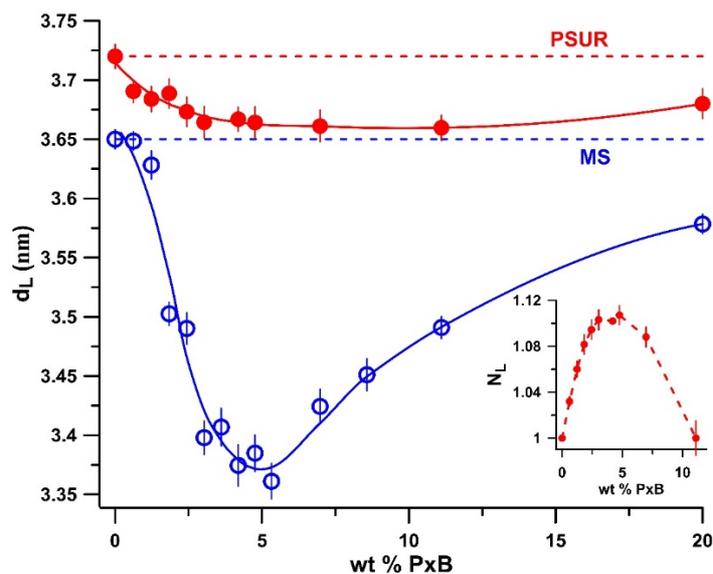

Fig. 4B Effect of polymyxin B on the thickness of the lipid bilayer $d_L$ of PSUR (red symbols) and MS (blue symbols). The dashed line displays the value of $d_L$ without the additive. Inset: The population of oligolamellar vesicles formed by the interaction of PxB with the ULVs of PSUR.

Figure 4A shows representative SANS curves of the PSUR mixture (black symbols); unilamellar PSUR vesicles (violet symbols); and unilamellar PSUR vesicles incubated with PxB at 37 °C. The SANS curve of the PSUR suspension hydrated with $D_2O$ (black symbols) shows a small peak at $q \sim 0.7$ nm$^{-1}$ reflecting the presence of OLV or MLV. The obtained repeat distance, $d \sim 9$ nm, indicates the first-order peak of a lamellar phase, consisting of SAXS data shown in Fig. 3A (red curve). The same figure (Fig. 4A, violet symbols) shows a typical scattering curve of polydisperse ULVs that were prepared by extrusion of the PSUR suspension, as described in Materials and Methods. The experimental curve $I(q)$ versus $q$ was fitted by the polydisperse unilamellar vesicle model with the aim of determining the thickness $d_L$. The fitting was processed with SasView software employing a built-in vesicle model[46] (SANS analysis, Fig. S9). The full violet line in Fig. 4A represents the best fit of the PSUR experimental data with $d_L = 3.72\pm0.02$ nm. Similarly to SAXS, SANS experiments were also performed repeatedly preparing samples from the non-expired drug of different batches. The thickness of the lipid bilayer, $d_L$, determined by the fitting (using the same model), ranges from 3.69 to 3.84 nm ($\pm0.02$ nm), with the average value $3.74\pm0.02$ nm. The value corresponds to the bilayer thickness reported $d_L = 3.76$ nm of bovine lung surfactant in fluid state[82], and $d_L = 4.13\pm0.02$ nm derived for Curosurf® at 40 °C by Filho et al.[59] For protein-free MS, $d_L = 3.65\pm0.05$ nm was obtained from the best fit of the SANS data at 37 °C using the same model (Fig. S11). The model was previously tested and, for comparison, we have obtained $d_L = 3.61 - 3.87$ ($\pm0.02$) nm for ULV from DPPC and DOPC in the fluid state, while $d_L = 4.20\pm0.02$ nm for DPPC vesicles in the gel state[91,92].

Detailed inspection of the SANS curves of the PxB-PSUR and PxB-MS mixtures (Figs. 4A and Fig. S11, S12) revealed that the addition of PxB induces the aggregation of EPS unilamellar vesicles. The effect is proved by the Kratky-Porod plot, $ln(I.q^2)$ vs. $q^2$, where $I(q)$ is the scattered intensity, and is shown for selected data in the inset in Fig. 4A. The analysis is based on the Guinier approximation[93,94] (SAXS analysis, Eq. S4). In this approach, the Kratky-Porod plot is linear for noninteracting particles (like ULVs), while possible correlations between objects such as an oligo- or multilamellar vesicle fail to meet the linearity [80]. Note that the plot of unilamellar PSUR vesicles (violet symbols) is linear (r$^2 \sim 0.999$), contrary to PSUR with 4.8 wt% of PxB (red circles). Instead, for PSUR with 11.1 wt% PxB (green triangles), the dependence is linear again

($r^2 \sim 0.999$). In SANS data analysis, any minor aggregation or presence of OLV is a hindrance to fitting experimental $I(q)$ data by using the vesicle model. Experimental $I(q)$ of the PxB – PSUR mixtures was fitted by using a paracrystal lamellar model[47] with respect to the lamella thickness ($d_L$) distribution and the number of lamellae in a single cluster. The stacks of lamellae formed by lipid bilayers separated by water layers are treated as a paracrystal to account for the repeating spacing ($d$). The parameters of the model[47] are: scale factor $\Gamma_m$, bilayer thickness $d_L$ (discussed in the next paragraph), number of layers $N_L$, and repeat distance $d$ (SANS analysis, Eqs. S5-S7). Our previous studies confirmed the suitability of the paracrystal lamellar model for lipid - additive oligolamellar vesicles (OLV)[91,95].

PxB is associated with negatively charged EPS vesicles (PSUR and MS) as a result of electrostatic forces, resulting in formation of OLV. The fitting output $N_L$, characterizing the number of interacting bilayers, is described in the lamellar paracrystal model by $Z_N(q)$ in the formula (Eq. S4). The inset of Fig. 4B shows the graph of $N_L$ as a function of PxB (wt%) obtained from the fitting of PxB-PSUR vesicles. $N_L$ shows a quasiparabolic dependence on the PxB content in the mixture. Note that for unilamellar PSUR vesicles, $N_L = 1$. $N_L$ values increase with PxB, reaching a maximum $N_L=1.11\pm0.01$ at ~ 5 wt% of PxB. According to Eq. S5, it means 89±1 % of unilamellar vesicles (ULV, $N_L=1$) and 11±1 % of oligo-lamellar objects with 2 bilayers on average. A further increase in the PxB content resulted in the recharge of the vesicle surface. We obtain $N_L = 1.00\pm0.01$ for PxB-PSUR with 11.1 wt% of PxB, showing that ULVs are dominant. It correlates well with the linearity of the Kratky – Porod plot in Fig. 3A, inset (green triangles). The repeat distance $d$ obtained from SANS correlates well with $d$ derived from the SAXS data of PxB-PSUR (Fig. S9). In the SANS model, the repeat distance is described by a Gaussian polydispersity with a width of ~ 12 – 15 %. The gray error bars in Fig. S13 are related to the polydispersity and indicate possible fluctuations of the repeat distance in OLV.

The plot of $N_L$ vs. wt% PxB for protein-free MS is shown in Fig. S11, inset. The dependence is similar, however, the fitting indicates a higher population of OLVs with 2 bilayers on average, ranged between 11 and 14 % ($N_L \sim 1.11 – 1.14$) in MS with 3-7 wt% of PxB. And OLV (~ 5 %) was detected even in the mixture with 20 wt% of PxB. The fitting indicated the larger spacing, the values of $d$ fluctuated from ~ 8.2 nm to ~ 7 nm with an increase in the content of PxB.

Interestingly, PxB molecules mediated "close mutual contact" of EPS bilayers even if they constitute ULV that are diluted enough to avoid any interaction between particles (1 wt% of the lipid). This finding correlates well with the concept that PxB mimics the properties of SP-B protein[38], and the formation of stable vesicle - vesicle contacts allowing the exchange of phospholipids[17,19]. A similar "condensing" effect was recently reported for the interaction of cationic nanoparticles and polymers with PSUR[77,96]; cationic liposomes with polyelectrolytes of opposite charges, such as DNA[97–99]; or in the context of antimicrobial mechanisms of cationic peptides that interact with bacterial model membranes[100]. However, it is worth stressing two points: 1) our experiments show that binding of PxB molecules onto EPS bilayers results in charge inversion of the membrane, as shown by the zeta potential and proved by SANS measurements. Therefore, great care must be taken in the potential administration to assess the amount of PxB mixed with PSUR to preserve the functionality of the EPS (< 5 wt% of PxB relative to the mass of the PSUR). 2) Note that SANS curves do not show any obvious sign of a strong aggregation of PxB - PSUR. Indeed, we did not observe massive precipitation due to the charge compensation in PxB – PSUR mixtures, as reported, for example, in studies of lipoplexes[101,102] (complexes of cationic vesicles – DNA) and also detected in the interaction of antimicrobial peptides with negatively charged bacterial model membranes[100]. It is a great benefit for the potential administration of PSUR enriched with moderate amounts of PxB that the galenic formulation is preserved, as it is essential for the rapid homogeneous spread of the drug in the lung alveoli[27,28,103].

*The effect of PxB on the lipid bilayer*
Fig. 3B summarizes the effect of PxB on the thickness of the lipid bilayer $d_L$ in both systems (PSUR and MS). Dashed lines mark the thickness of the lipid bilayer in the unilamellar vesicles of PSUR and MS, respectively. There is an apparent thinning of the lipid membrane in both systems due to PxB binding. The effect is negligible for PSUR. Instead, there is a visible effect of PxB on the bilayer thickness $d_L$ of protein-free MS vesicles (Fig. 3B, empty blue symbols). We observe a thinning reaching a minimum value of $d_L$ at ~ 5 - 6 wt% of the antibiotic, that is, in the vicinity of electroneutrality of PxB-EPS (Fig. 1). The further increase of the PxB content induces a consecutive increase in $d_L$.

To gain further insight into PxB binding, we monitored changes in lateral pressure in PSUR and MS after interaction with PxB by excimer fluorescence spectroscopy in two regions of the lipid bilayer: near the hydrophilic/hydrophobic interface (Pyr4PC) and near the hydrophobic core (Pyr10PC)[54]. Typical fluorescence emission spectra of both probes are shown and characterized in Fig. S14 (Lateral pressure, Supporting Information). The interfacial energy concentrated over the small thickness of the lipid membrane, relative to its lateral size, is balanced by high magnitudes internal pressures[104]. By definition, positive lateral pressure corresponds to repulsive interactions and negative pressure corresponds to attractive interactions. The sum of all interactions along the normal of the membrane is equal to zero in a tensionless lipid bilayer in water. It follows that if an additional molecule interacts with the membrane, an array of tension (contracting) and pressure (expanding) components alter the lateral pressure distribution across the membrane. Fig. 5 shows changes in the lateral pressure of the PSUR (red symbols) and MS (blue symbols) bilayer monitored by excimer fluorescence spectroscopy at two depths of the hydrophobic core of the lipid bilayer (Pyr4PC and Pyr10PC) at 45 °C.

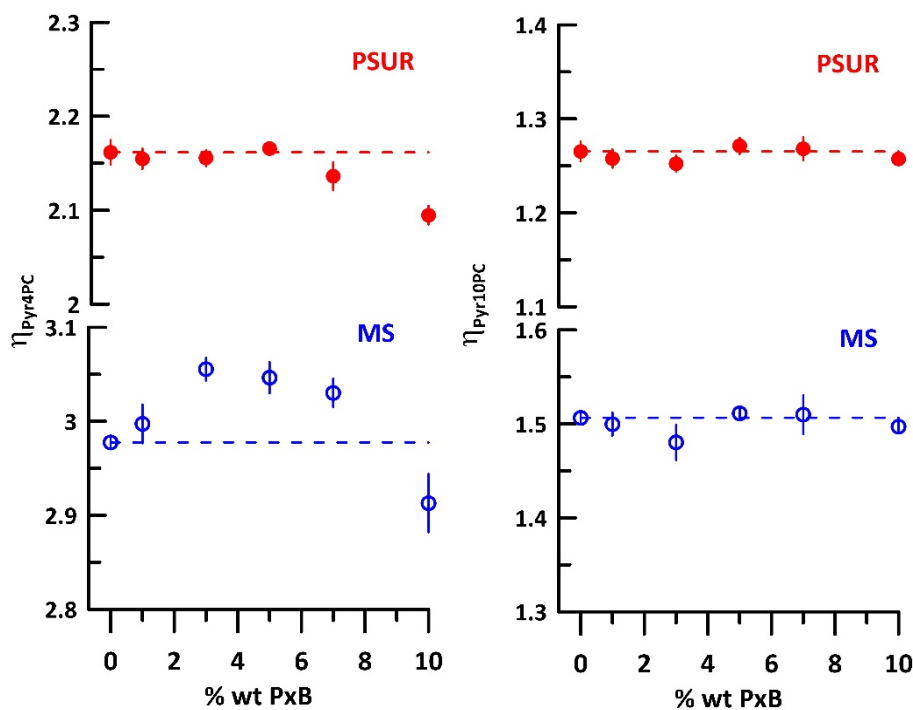

Fig. 5 Effect of PxB on the lateral pressure η of lipid bilayer PSUR (red symbols) and MS (blue symbols) derived from emission spectra of Pyr4PC (left panel) and Pyr10PC (right panel) fluorescent probes (at 45 °C). Error bars denote the standard deviation from the time-based measurements of fluorescence intensities.

The values obtained with Pyr10PC are lower than those obtained with Pyr4PC, consistent with a lower lateral pressure in the hydrophobic core than in the hydrophilic/hydrophobic interface of the bilayer. For bilayers without PxB η = 2.16 and

1.27 for PSUR and η = 2.98 and 1.51 for MS were obtained with Pyr4PC and Pyr10PC, respectively. Therefore, overall lateral pressures are lower in PSUR than in the MS bilayer, although markedly higher than in a single component fluid dioleoylphosphatidylcholine (DOPC) bilayer[105].

The addition of PxB does not affect the lateral pressure deeper in the hydrophobic region of the bilayer, monitored by Pyr10PC. In both systems, the $\eta_{Pyr10PC}$ values oscillate around the value of η detected for bilayers without PxB (marked by dashed lines in Fig. 5). However, we observed differences in the lateral pressure between PSUR and MS close to the hydrophilic/hydrophobic interface (monitored by Pyr4PC). In PSUR, the lateral pressure does not change with increasing PxB content up to ~ 5 wt%, and decreases slightly above this content. In PxB-MS, the lateral pressure close to the polar headgroup (monitored by Pyr4PC) increases with the PxB content up to a break point (~ 5 wt%) when PxB compensates charges of the MS. In the same concentration range of PxB, the DSC data show an increase in *Tm* (Fig. 2B). The results of both methods demonstrate denser packing of the lipid acyl chains in the PxB-MS than in MS alone. The obtained results are consistent and lead us to propose the interaction mode: The presence of uncompensated charges at the bilayer surface induces repulsion between individual polar fragments of neighboring molecules, resulting in the lateral expansion of the bilayer (and consecutively to the long-range disorder, as discussed in the previous paragraph). The adsorption of PxB onto the negatively charged surface of the MS bilayers is driven electrostatically. Moreover, because of the hydrophobic effect, the short hydrophobic tail of PxB penetrates into the hydrophobic region of the MS lipid bilayer. Due to the chain length mismatch, the PxB acyl chain creates voids in the hydrophobic region of the bilayer, which are compensated for by bending the chains or the trans−gauche isomerization of hydrocarbon chains, resulting in the thinning of the lipid bilayer (as previously shown[106,107]). Note that the unsaturated acyl chains of MS are composed of oleic acid (18:1) and linoleic acid (18:2) with the double bond position in carbon C9 (and C12 for 18:2), while the hydrophobic tail of PxB is shorter, formed by 6-methyloctanoic acid (polymyxin B1) or 6-methylheptanoic acid (polymyxin B2)[2]. Therefore, the increase in lateral pressure close to the hydrophobic-hydrophilic interface, monitored by Pyr4PC, reflects the denser acyl chains packing resulting from intercalated PxB tails between acyl chains of lipids. This, in turn, leads to an increase in the temperature of the gel-fluid phase transition. A denser lipid pack induced by PxB was previously reported[14,108]. Electrostatic attraction is dominant up to PxB content ~ 5-7 wt%, when zeta potential indicates electroneutrality. PxB – MS mixtures with compensated charges are manifested by a well-ordered stack of lamellae, and, we detected an aggregation of MS unilamellar vesicles into OLV. Interestingly, for PxB > 6 wt% $d_L$ increases to $d_L$ = 3.59 nm for MS in the presence of 20 wt% of the drug (Fig. 4B). An entropically driven hydrophobic effect to minimize unfavorable contact of the PxB lipophilic acyl chain with water secures additional partitioning of PxB molecules into a noncharged bilayer. Finally, this results in the inversion of the PxB-MS surface charge. In fact, the zeta potential (Fig. 1) reaches positive values, although not exceeding + 5 mV in mixtures with 10 wt% < PxB ≤ 20 wt%. Repulsion between positively charged segments of neighboring PxB molecules induces disorder in the polar part of the lipid bilayer, allowing deeper penetration of water molecules. An excess of positive charges causes lamellae repulsion and fluctuations, leading to changes in the bilayer curvature due to the break of the OLV back to the unbound ULV. We assume that a small drop in lateral pressure at 10 wt% of PxB, monitored by Pyr4PC, reflects these changes. Therefore, the increase in $d_L$ ~ + 0.2 nm detected in PxB-MS mixtures at PxB > 6 wt% (Fig. 4B) may result from both the steric thickness of the lipid bilayer affected by the adsorption of PxB, and the fluctuations of the lamellae that increase the bound – unbound transitions[109]. We did not detect an increase in the thickness of the lipid bilayer in the range 5 wt% ≤ PxB ≤ 10 wt% when aligned stacks of lipid bilayers of the PxB-MS are deposited on silicon wafer and hydrated by vapor (99 RH%) (ongoing research of the team).

The effects of PxB on the lipid bilayer are similar in the two EPS. However, in PSUR, the effects are "damped", except for electrostatic attraction, which modulates the repeat distance $d$ and the long-range order (Fig. 3). We conclude that PxB molecules do not disturb the hydrophobic region of the PSUR lipid bilayer. The minor thinning of $d_L$ can be attributed to structural changes in the polar part of the bilayer accommodating PxB molecules, their effect on the elasticity and curvature of bilayers undergoing transformation from ULV to OLV, and their reversal to ULV with increasing content of the drug.

Recently, Khondker et al.[14] have shown that membrane charge and lipid packing play a crucial role in PxB binding, its penetration into the membrane and eventually its damage. Electrostatic interaction between PxB and the membrane surface enhances the penetration depth of PxB and reduces the tilt of PxB within the bilayer; however, an increase in the membrane packing and rigidity can inhibit the insertion of PxB into the bilayer. On the basis of electron density profiles constructed from SAXD data of highly oriented multilamellar stacks of lipid bilayers and molecular dynamic simulations, the authors proved different orientations of PxB molecules: they were found to lie flat on membranes with saturated acyl chains up to inserted parallel to the normal on a membrane rich in lipids with unsaturated acyl chains. The latter orientation of PxB can cause membrane damage. Kinetic measurements have shown rapid insertion of the hydrophobic tail of PxB into the POPC (C16:0-C18:1PC) membrane, and its damage was consecutively monitored by leakage of the encapsulated fluorescent probe[14,110]. Our experimental findings are in line with the proposed model. In protein-free MS, unsaturated acyl chain lipids represent ~ 50 mol% and all results indicate insertion of PxB molecules into the bilayer up to their content ~ 5-7 wt% (~ 30 – 21 molecules of lipid per PxB, derived from the composition). The drug at its higher content is adsorbed onto the bilayer, resulting in the positively charged surface. We detected changes in the thickness of the lipid bilayer of MS; however, no damage was observed to the bilayer. Interestingly, for PSUR, all of the obtained results indicate that PxB molecules lie flat on the lipid bilayer and manifest themselves only by electrostatic interactions. Let us inspect the difference between the two systems: The electrostatics of binding of PxB onto the negatively charged lipid bilayer are very similar for both systems studied, as discussed in the previous paragraph. After binding, the depth of PxB insertion into the lipid bilayer depends on its rigidity and thus the bilayer composition. We focus on the rigidity of the bilayer. Due to the complex composition of natural PSUR (Table S1), we infer the penetration effect of PxB from the ratio of saturated and unsaturated fatty acids (sFA and uFA) that form the zwitterionic PC. PxB was found to damage the POPC bilayer[14], therefore POPC for which sFA/uFA = 1 can serve as an indicator of the bilayer damage. Rudiger et al.[111] reported the acyl chain composition for PC and PE of Curosurf® (PSUR). Based on the data in Table S2, the ratio sFA/uFA ~ 3.6 for the protein-free MS system, while sFA/uFA ~ 3 for the PSUR bilayer. The higher value of sFA/uFA indicates better protection against penetration of PxB. Thus, the composition of phosphatidylcholines (PC) only indicates a lower protection of PSUR against penetration of PxB when compared with MS. The fraction of phosphatidylethanolamines (PE) represents 4.5 – 7.5 wt% of the total mass of PSUR (Table S1). The ratio sFA/uFA ~ 0.4 expressed in the fatty acid composition of PEs (Table S2) indicates loosely packed molecules if considered "self-reliantly". However, it was proved that the conformational order of acyl chains in PC bilayers increases markedly with the addition of PE[112–114]. Therefore, tighter packing of the acyl chains in mixed bilayers can increase lateral order, supported by the complementarity of the "shapes" of the PE and PC molecules. The effective cross-sectional area per PE headgroup is smaller than that of acyl chains[115]. Hence, the relief from "packing frustration" of PE in the presence of PC with its bulky headgroup may be one of the mechanisms that support the formation of microdomains, even regular, superlattice-like arrangements of polar headgroups in PE/PC bilayers[116]. Thus, PEs, even in small amounts, contribute to the PSUR bilayer rigidity. Similarly, sphingomyelins (SM) contribute to the rigidity of the PSUR. Its content was found to vary from 1.8 to 8 wt% (Table S1) in Curosurf®, and the detected composition SMs with sphingoid base d18:1/n-acyl with n-acyl of 16:0, 22:0, 24:0 and 24:1[59]. Generally, the phase transition temperature *Tm* of SMs is much higher compared to that of PC analogues because of the difference in the *cis* double bond position in the chain. Due to their chain mismatch, they induce lateral heterogeneity and immiscibility generating domains[117]. Finally, we should also consider the contribution of the surfactant specific proteins, SP-B and SP-C. Their amount does not exceed 1.5 wt%. However, their properties play an indispensable role in the functionality of PSUR[40,118]. Both proteins are cationic, thus, their interactions with the negatively charged components of PSUR are crucial. Both proteins slightly increase the order of lipids in the fluid phase without having any apparent effect on the gel phase[119], and support the formation of lateral domains, although their location in the bilayer is different[58]. A recent study confirms the SP-B induced model for membrane aggregation[120]. A shallow anchoring of SP-B onto the lipid bilayer without a deep perturbation of the acyl chains[58,121] is in line with the proposed orientation of PxB molecules lying flat on the PSUR surface. The cationic SP-B that preferentially interacts with phosphatidylglycerol[116] lowers the negative surface charge of PSUR and, along with it, forms a barrier for penetration of the PxB tail. In summary, all "minor" components incerase the resistance of the PSUR lipid bilayer to its perturbation by PxB molecules, while profiting from the electrostatic interactions, maintaining the integrity of PSUR. In other words, enrichment of PSUR with a reasonable amount of PxB reinforces the role of the SP-B protein. Finally, this is in agreement with the previously reported ability of PxB to mimic the properties of SP-B protein[38].

As we have seen, the values of $\eta_{Pyr10PC}$ just oscillate around the value of the lateral pressure obtained for the system without any additive. Thus, the PxB molecules do not disturb the central part of the lipid bilayer in any of the two systems. In both lipid mixtures, $\eta$ values are higher than with those detected in the bilayer prepared from unsaturated dioleoyphosphatidylcholine (diC18:1PC)[105], or diC18:1PC/diC18:1PE mixture[54]. We hypothesize that the higher lateral pressure could reflect the domain-like character of both mixtures because of the nonideal miscibility of the compounds. However, it should be stressed that additional experimental methods are necessary to shed more light on the lateral ordering of both systems studied. Surprisingly, $\eta$ is significantly higher for the protein-free MS, the four phospholipid mixture, than for PSUR composed of many chemical species. This fact underscores the "cleverness" of the nature, revealing that the composition of lipids is an intrinsic property of each type of cell and its synthesis is genetically controlled to maintain its function.

**Conclusion**

We are certain that exogenous pulmonary surfactant (EPS) as a vehicle for various drugs and other compounds is a promising approach in a variety of circumstances requiring pulmonary delivery. EPS, because of its rapid spreading properties, is capable of transporting pharmaceutical agents to the terminal air spaces. The premise of surfactant-based drug administration would limit potential side effects associated with systemic drug administration. However, this requires careful inspection of the mutual interactions between EPS and the transported agent. This study inspects one of these therapeutic possibilities, to assess the effect of PxB, an antibiotic targeting Gram-negative bacteria, on the structure and thermodynamics of porcine Curosurf® exogenous pulmonary surfactant (PSUR). The PxB – PSUR system was studied by several techniques in order to unravel the mode of interaction. We found that PxB does not affect the thermal behavior of PSUR, and does not change the gel- to fluid phase transition temperature ($Tm$). PxB affects slightly the thickness of the lipid bilayer ($d_L$) of PSUR; however, the hydrophobic tail of PxB does not intrude into the bilayer. This conclusion is supported by measurements of the lateral pressure at two depths of the hydrophobic region of the bilayer employing the excimere fluorescence technique. The obtained results are consistent and lead us to conclude that cationic PxB molecules lie on the bilayer surface of PSUR, strengthening the multilamellar structure of PSUR. In other words, enrichment of PSUR with a reasonable amount of PxB reinforces the role of the SP-B protein. Electrostatic interactions dominate in the PxB - PSUR system.

Conversely, binding of PxB to EPS bilayers can result in inversion of the surface charge of the membrane. Consequently, the amount of PxB must be carefully assessed (less than 5 wt% of the PSUR mass). A protein-free model system MS composed of four phospholipids mimics natural EPS well but is less resistant to PxB penetration. The parallel study of both systems revealed the protective role of the minor components of PSUR (PEs, SMs, SP-B, and SP-C proteins) against PxB penetration to its bilayer. So far, our studies did not find any results discouraging the concept of a combined treatment with PxB enriched Curosurf®.

**Supporting information:** Structure of polymyxin B, Tables of Curosurf® composition, DSC scans, SAXS/WAXS data and analysis, SANS data and analysis, excimer fluorescence spectra

**Acknowledgments:** SAXS experiments were performed at the BL11-NCD beamline at ALBA Synchrotron with the collaboration of ALBA staff. The authors thank ALBA staff for their help. D.U., N.K. and L.H. thank the staff of LLB CEA Saclay for the hospitality and granted beam time at the PAXY spectrometer. The research that led to these results was supported by grants VEGA 1/0223/20, 1/0305/24, APVV-17-0250, FaF/27/2023 and FaF/28/2023. This work benefited from the use of the SasView application, originally developed under the NSF award DMR-0520547. SasView contains code developed with funding from the European Union's Horizon 2020 research and innovation program under the SINE2020 project, grant agreement No 654000.

**List of abbreviations:** EPS, Exogenous pulmonary surfactant; PxB, Polymyxin B; PSUR, Poractant alpha - Curosurf®; MS, Model system of pulmonary surfactant; DSC, Differential scanning calorimetry; SAXS, Small-angle X-ray scattering; WAXS, Wide-angle X-ray scattering; SANS, Small-angle neutron scattering; Tm, Gel-to-liquid phase transition temperature; Dab, Diaminobutyric acid; LPS, Lipopolysaccharide; ULV,OLV,MLV, Uni-, oligo-, multilamellar vesicles; SP-B, SP-C, Pulmonary surfactant specific protein B and C; DPPC, 1,2-dipalmitoyl-phosphatidylcholine (diC16:0PC); POPC, 1-palmitoyl-2-oleoyl-phosphocholine (16:0-18:1PC); PLPC,1-palmitoyl-2-linoleoyl-phosphocholine (16:0-18:2PC); POPG, 1-palmitoyl-2-oleoyl-phosphotidylglycerol (16:0-18:1PG); Pyr4PC, Pyr10PC, 1,2-bis-pyrenebutanoyl-, 1,2-bis-pyrenedecanoyl phosphatidylcholine; PL, Proteolipidic mixture; ΔH, Enthalpy; d, Repeat distance; $d_W$, Thickness of water layer; $d_L$, Thickness of the lipid bilayer; q, Momentum transfer, q-vector; λ, Wavelength; 2θ, Scattering angle; ρ, Neutron scattering length density (NSLD); PC, Phosphatidylcholine; η, Lateral pressure; PG, Phosphatidylglycerol; PI, Phosphatidylinositol; FWHM, Full width at half maximum; $L_β$, Gel phase; Lα, Liquid-crystalline phase; cp, Heat capacity; T, Temperature; $N_L$, number of layers; $Γ_m$, Scale factor; DOPC, Dioleoylphosphatidylcholine; SAXD, Small-angle X-ray diffraction; sFA, Saturated fatty acids; uFA, Unsaturated fatty acids; PE, Phosphatidylethanolamine; SM, Sphingomyelin